\documentclass[aps,pre,twocolumn,groupedaddress,showpacs]{revtex4}

\usepackage{graphicx}
\usepackage{amsmath}

\begin{document}

\title{Staggered repulsion of transmission eigenvalues \\in
symmetric open mesoscopic systems}

\author{Marten Kopp}
\author{Henning Schomerus}
\affiliation{Department of Physics, Lancaster University,
Lancaster, LA1 4YB, United Kingdom}

\author{Stefan Rotter}
\affiliation{Department of Applied Physics, Yale University, New
Haven, CT 06520, USA}

\date{\today}

\begin{abstract}
Quantum systems with discrete symmetries can usually be desymmetrized, but
this strategy fails when considering transport in open systems with a
symmetry that maps different openings onto each other. We investigate the
joint probability density of transmission eigenvalues for such systems in
random-matrix theory. In the orthogonal symmetry class we show that the
eigenvalue statistics manifests level repulsion between only every second
transmission eigenvalue. This finds its natural statistical interpretation as
a {\em staggered} superposition of two eigenvalue sequences. For a large
number of channels, the statistics for a system with a lead-transposing
symmetry approaches that of a superposition of two uncorrelated sets of
eigenvalues as in systems with a lead-preserving symmetry (which can be
desymmetrized). These predictions are confirmed by numerical computations of
the transmission-eigenvalue spacing distribution for quantum billiards and
for the open kicked rotator.
\end{abstract}


\maketitle

\section{Introduction}

Mesoscopic systems exhibit variations in their phase-coherent electronic
transport properties that are conveniently characterized via statistical
approaches. Geometries that classically give rise to chaotic motion typically
display universal fluctuations which can be captured using ensembles of
random scattering matrices \cite{B97}. For normal conductors the universal
properties fall into Dyson's three universality classes with symmetry index
$\beta=1,2,4$ \cite{D62}, while a further seven universality classes can be
identified in the presence of superconducting  or chiral particle-hole
symmetries \cite{AZ97}. A powerful tool to distinguish these ensembles is the
amount of level repulsion between the transmission eigenvalues $T_n$. These
eigenvalues determine fundamental transport properties such as the
conductance $G$  or the shot-noise Fano factor  $F$ \cite{B97,BB00}. In the
Dyson ensembles, the probability density to find two closely spaced adjacent
transmission eigenvalues with small distance $s=T_{n+1}-T_{n}$ is suppressed
as $P(s)\propto s^\beta$ \cite{mehta-2004,haake-2001}. This introduces a
stiffness in the transmission-eigenvalue sequence which suppresses the
fluctuations of the conductance and of the Fano factor when compared to the
case of uncorrelated transmission eigenvalues (the latter being
characteristic for classically integrable systems with a complete set of good
quantum numbers) \cite{B97,BB00}.

From the investigation of closed systems it is well known that
discrete symmetries result in a reduction of level repulsion. In
such systems, desymmetrization delivers independent variants of
the system which differ by the boundary conditions on the symmetry
lines (e.g., Dirichlet and Neumann boundary conditions for
eigenfunctions of odd and even parity, respectively). The
statistics of the desymmetrized versions can depend on the
dimensionality of the irreducible representation
\cite{keating:robbins}, but still remain within the conventional
universality classes. The combined level statistics is then built
by superimposing the independent level sequences of the
desymmetrized variants \cite{mehta-2004}. In open systems, this
concept of desymmetrization can be directly applied as long as the symmetry in question preserves the shape and position of the leads \cite{GMMB96, BM96}.

This paper is motivated by the observation that systems with a
lead-transposing symmetry (which maps {\em different} openings
onto each other while leaving the dynamics in the system
unchanged) exhibit transport properties that can only be
understood as collective features of the desymmetrized variants of
the system \cite{GMMB96, BM96}. An obvious indication of this
complication is the fact that the symmetry-reduced variants only
possess a reduced number of leads (we concentrate on systems with
two leads, for which the desymmetrized variants only possess a
single lead). We demonstrate that
such systems exhibit nevertheless
a reduced repulsion of transmission eigenvalues which is similar
to that for
systems with
a lead-preserving symmetry. For a large number $N$ of transport
channels, the local statistical fluctuations in the
eigenvalue
 sequence
indeed become indistinguishable for both types of symmetry.
However, for a small numbers of channels, the statistics differ
from each other,
which can be traced back to the absence or presence of $1/N$
corrections in these ensembles.

In the specific case
of
$\beta=1$, we derive exact closed expressions
for the joint probability density of transmission eigenvalues
thereby gaining detailed insight into these statistical features.
In particular, we find
for both the lead-preserving and the lead-transposing symmetry class
that level repulsion occurs only between every second transmission
eigenvalue. The fluctuations in the transmission eigenvalue
sequence hence find their most natural statistical interpretation
in a staggered superposition of two independent level
 sequences.
In such a superposition, the transmission eigenvalues alternate
between the two sequences when they are ordered by magnitude.

The exact expressions for the joint probability density with
$\beta=1$
are different for the two
types of symmetry. Hence, the details of the transport statistics
for a lead-transposing symmetry deviate from those for a
lead-preserving symmetry. We show that these deviations are most
significant for a small number of channels, while for a large
number of channels the local
eigenvalue statistics do indeed converge
onto each other.

Previous studies of open systems with lead-transposing or lead-preserving
symmetries have derived the distribution of transmission eigenvalues for one
or two open channels and the one-point density for arbitrary numbers of
channels \cite{GMMB96,BM96,SSML05,gopar-2006-73}. For time-reversal symmetric
systems with $\beta=1$, a key observation of these works was an enhancement
of universal fluctuations for both types of symmetry (when compared to
asymmetric systems). For systems with a lead-transposing symmetry it was
found that the weak localization correction is vanishing, leading to ensemble
averaged expressions for the conductance and for the shot noise Fano factor
which are entirely independent of the channel number N \cite{gopar-2006-73}.
The underlying staggered level statistics embodied in the joint distribution
of transmission eigenvalues provides a unifying explanation for all of these
observations. We verify our predictions by numerical computations for quantum
billiards \cite{RTWTB00,RAB07} and for the open kicked rotator \cite{TTSB03,
JSB03, OKG03}.

This paper is organized as follows. Section \ref{sec:2} provides
background information
on
the scattering approach to transport and
on standard random-matrix theory.  In Sec.\ \ref{sec:2a} we revisit
the case of systems with a lead-preserving symmetry and provide
the exact reformulation of the eigenvalue statistics in the
orthogonal symmetry class ($\beta=1$) as a staggered superposition
of two eigenvalue sequences. Section \ref{sec:3} concerns systems with
a lead-transposing symmetry. In particular, for $\beta=1$ we
derive the exact joint probability density of transmission
eigenvalues for arbitrary $N$, and show that this again takes the
form of a staggered eigenvalue sequence. We also describe the
convergence of the local statistics for both types of symmetry,
which emerges in the limit $N\to\infty$. Section \ref{sec:4}
provides numerical results that illustrate the similarities and
differences of the random-matrix ensembles
for the two symmetry classes.
This section also contains the comparison to specific model
systems. Section \ref{sec:5} provides a summary and discussion of
our main results.

\section{Basic Concepts\label{sec:2}}

\begin{figure}[t]
\includegraphics[width=0.8\columnwidth, keepaspectratio]{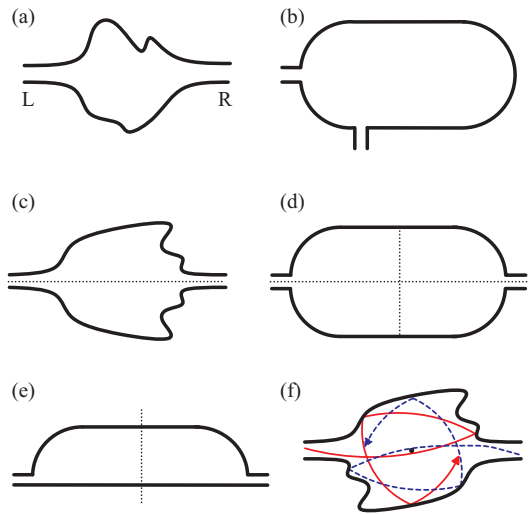}
\caption{(Color online) Sketches of quantum billiards (a,b)
without any spatial symmetry, (c) with a lead-preserving symmetry,
(d) with both a lead-preserving as well as a lead-transposing
symmetry, (e) with a lead-transposing reflection symmetry, and (f)
with  a lead-transposing inversion symmetry. The inversion
symmetry in panel
(f)
survives in the presence of a finite magnetic field, as is
indicated by a symmetric pair of trajectories.}
\label{fig1}
\end{figure}

\subsection{Scattering approach to transport}

Figure \ref{fig1} depicts open two-dimensional quantum billiards representing
mesoscopic systems with two attached leads (L -- left and R -- right), each
carrying $N$ incoming and $N$ outgoing modes. The systems in Fig.\
\ref{fig1}(a,b) are asymmetric while those in Fig.\ \ref{fig1}(e,f) possess a
lead-transposing reflection or inversion symmetry, respectively. The
inversion symmetry in panel (f) survives in the presence of a finite magnetic
field, which however breaks time-reversal symmetry (an inversion symmetry
does not induce a generalized antiunitary symmetry, in contrast to reflection
symmetries \cite{haake-2001}). In the middle panels, Fig.\ \ref{fig1}(c)
shows a system with a lead-preserving symmetry, and Fig.\ \ref{fig1}(d) shows
a system which possesses both a lead-preserving (up-down) and a
lead-transposing (right-left) symmetry.

In order to describe the phase-coherent transport through these
systems for small bias voltage $V$, one solves  the
Schr{\"o}dinger equation for fixed values of the $2N$ amplitudes
${\bf a}=[a_n^{(L)},a_n^{(R)}]^T$ in the incoming modes.  This
results in linear relations ${\bf b}=S{\bf a}$ for the $2N$
amplitudes ${\bf b}=[b_n^{(L)},b_n^{(R)}]^T$ in the outgoing
modes, which delivers a $2N\times 2N$-dimensional scattering
matrix of the form
\begin{equation}
S=\left(\begin{array}{cc}r&t'\\t&r'\end{array}\right).
\label{eqn:srt}
\end{equation}
Here $r, r', t, t'$ are $N\times N$-dimensional matrices
describing reflection at each lead and transmission from one
lead to the other, respectively.

The scattering matrix is unitary, and its structure is further
constrained by symmetries of the system.  The three main
universality classes arise for systems with time-reversal and
spin-rotation symmetry (orthogonal symmetry class with $S=S^T$,
symmetry index $\beta=1$), systems without time-reversal symmetry
(unitary symmetry class with no constraints on $S$, $\beta=2$),
and systems with time-reversal but broken spin-rotation symmetry
(symplectic symmetry class composed of self-dual matrices $S=S^R$,
$\beta=4$). Spatial symmetries entail additional constraints on
the scattering matrix, which are detailed in Secs.\ \ref{sec:2a}
and \ref{sec:3}.

The transmission eigenvalues $T_n$ are defined as the
eigenvalues of the hermitian matrix $tt^\dagger$. In the case
of spin-independent transport or Kramers degeneracy (the latter
occurs for $\beta=4$), the transmission eigenvalues are twofold
degenerate. We then only account for each pair of eigenvalues
once and introduce a spin-degeneracy factor $\alpha=2$. When
the two-fold degeneracy is lifted then $\alpha=1$. From here
on, $N$ refers to the number of distinct transmission
eigenvalues (ignoring accidental degeneracies). Furthermore we
will assume that the transmission eigenvalues are ordered by
magnitude,
\begin{equation}
T_1\leq T_2\leq T_3\leq\ldots\leq T_N,
\label{eq:ordering}
\end{equation}
as this results in a number of technical simplifications. The
conductance quantum is defined as $G_0=\alpha e^2/h$.

With these conventions, the transmission eigenvalues
determine fundamental transport properties such as the
conductance via
\begin{equation}
 G=G_0\sum_{n=1}^NT_n
\end{equation} and the shot-noise power via
\begin{equation}
P=2 G_0eV\sum_{n=1}^N T_n(1-T_n).
\end{equation}
Here $V$ is the bias voltage, which is assumed to be small.

\subsection{Dyson's circular ensembles}

Random-matrix theory delivers a statistical description of
transport by drawing the scattering matrices from ensembles of
unitary matrices which obey the constraints of the given
universality class. For the three main universality classes
with $\beta=1$, $2$, or $4$, random-matrix theory is based on
Dyson's circular ensembles, for which the probability measure
is given by the Haar measure of unitary symmetric, unitary, or
unitary self-dual matrices, respectively. The joint probability
density of transmission eigenvalues then takes the form
\cite{B97}
\begin{equation}
P(\{T_n\})
\propto \prod_{m>n} \left(T_m-T_n\right)^\beta \prod_l
T_l^{-1+\beta/2}.\label{eq:dyson}
\end{equation}

The first product in Eq.\ (\ref{eq:dyson}) involves  pairs of
transmission eigenvalues and favors sequences in which
neighboring transmission eigenvalues do not approach each other
closely. (As we have ordered the transmission eigenvalues by
magnitude, all differences $T_m-T_n$ are positive.) This
suppresses fluctuations in the eigenvalue sequence and ultimately
results in conductance fluctuations of the order of a single
conductance quantum, which for large $N$ approach the asymptotic
value
\begin{equation}\mbox{var}\,G/G_0=\frac{1}{8\beta}.
\label{eq:varg}
\end{equation}

For large $N$, the one-point probability density of transmission
eigenvalues approaches
\begin{equation}\label{eq:onepoint}
P(T)=\frac{1}{\pi\sqrt{T(1-T)}}.
\end{equation}
The second product in Eq.\ (\ref{eq:dyson}) induces an asymmetry
into this bi-modal distribution, which for large $N$ results in
the weak-localization correction
\begin{equation}
\langle
G\rangle-\frac{N}{2}G_0=G_0\left(\frac{1}{4}-\frac{1}{2\beta}\right)
\label{eq:dg}
\end{equation}
of the ensemble-averaged conductance.

An insightful quantity derived from the joint probability density
$P(\{T_n\})$ is the distribution  $P(s)$ of spacings
$s=T_{n+1}-T_n$ between neighboring transmission eigenvalues. For
uncorrelated eigenvalues with average spacing ${\bar s}$ one would
expect a Poisson distribution,
\begin{equation}
P(s)={\bar s}^{-1}e^{-s/\bar s},\label{eq:poisson}
\end{equation}
while for the circular ensembles and $N\gg1$, the spacing
distribution can be well approximated by the Wigner distributions
\cite{mehta-2004,haake-2001},
\begin{equation}
P(s)=\left\{\begin{array}{ll}
\frac{\pi}{2{\bar s}^2}s \exp\left(-\frac{\pi s^2}{4{\bar s}^2}\right)&\beta=1\\
\frac{32}{\pi^2{\bar s}^3}s^2 \exp\left(-\frac{4 s^2 }{\pi{\bar s}^2}\right)&\beta=2\\
\frac{2^{18}}{3^6\pi^3{\bar s}^5}s^4 \exp\left(-\frac{64
s^2}{9\pi{\bar s}^2}\right)&\beta=4
\end{array}\right.
. \label{eqn:wignerdistribution}
\end{equation}

Lead-preserving and lead-transposing symmetries entail further
constraints on the scattering matrix. The consequences of these
constraints for the transmission eigenvalue statistics are explored in
the remainder of this paper.

\section{Lead-preserving symmetries\label{sec:2a}}

\begin{figure}[t]
\includegraphics[width=\columnwidth, keepaspectratio]{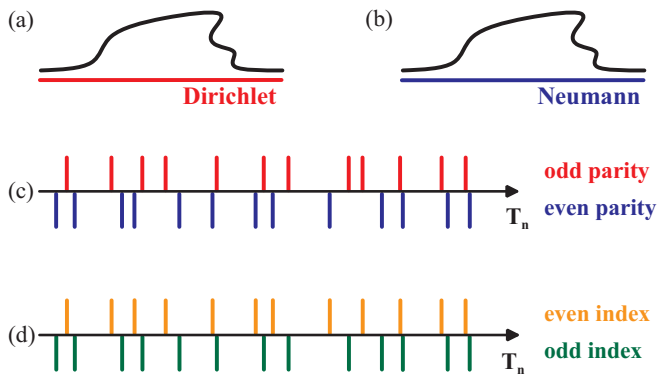}
\caption{(Color online) (a,b) Desymmetrization of the quantum
billiard with a lead-preserving reflection symmetry, shown in Fig.
\ref{fig1}(c). (c) Sketch of the individual transmission eigenvalue sequences of
fixed parity. (d)
Reorganization as a staggered level sequence, where transmission
eigenvalues alternate after ordering by magnitude.}
 \label{fig:leadpreserving}
\end{figure}

A useful reference point for our subsequent investigation of
systems with a lead-transposing symmetry (in Sec.\ \ref{sec:3})
are open systems with a lead-{\em preserving} symmetry, to which
one can directly apply the standard ideas of desymmetrization. The
goal of the present section is to reformulate the resulting
random-matrix statistics for
the case of a lead-preserving symmetry with
$\beta=1$ as a staggered level repulsion, as this will allow us to
establish a connection to the case of a lead-transposing symmetry.

\subsection{Constraints on the scattering matrix}

An example of a system with a lead-preserving reflection symmetry
is shown in Fig.\ \ref{fig1} (c). Figure \ref{fig:leadpreserving}
(a,b) shows the desymmetrized version of the system, which is
halved at the symmetry line. Dirichlet boundary conditions on the
line of symmetry select scattering wave functions with an odd
parity, while Neumann boundary conditions yield even parity.
Consequently, the transmission matrix $t$ assumes a block
structure where each block corresponds to a given parity. As
dictated by the one-dimensional transverse-mode quantization in
the leads, the block of even parity has dimension
$N_1\equiv[(N+1)/2]$, while the block of odd parity has dimension
$N_2\equiv[N/2]$ (here $[\cdot]$ denotes the integer part of a
number). Hence, both blocks have either the same size (when
$N=N_1+N_2$ is even), or the block with even parity is by one
larger than the block with odd parity (when $N$ is odd).

The total transmission-eigenvalue sequence is therefore obtained
from a  superposition of two sequences of size $N_1$ and $N_2$
[for illustration see Fig.\ \ref{fig:leadpreserving}(c)]. In
order to fix the way we address the elements of this
superposition, we impose the ordering of Eq.\ (\ref{eq:ordering})
and denote by ${\cal P}$ the set of all strictly increasing
sequences of indices $I_n\in\{1,2,3,\ldots,N\}$, where each
sequence is of length $N_1$. Such sequences are of the form
$I=(I_1,I_2,\ldots,I_{N_1})$, where $1\leq I_1< I_2< I_3<\ldots<
I_{N_1}\leq N$. For each
sequence
we also define a complementary sequence $\bar I=(\bar I_1,\bar
I_2,\bar I_3,\ldots,\bar I_{N_2})$, which consists of the indices
$1\leq \bar I_1< \bar I_2< \bar I_3<\ldots< \bar I_{N_2}\leq N$
not contained in $I$. This partition delivers two ordered
subsequences $T_{I_n}$ and $T_{\bar I_n}$.

\subsection{Conventional random-matrix theory}

Within random-matrix theory, the joint probability distribution of
the total transmission-eigenvalue sequence is the sum of the
corresponding probabilities for each way to distribute the
transmission eigenvalues into two sets containing $N_1$ and $N_2$
eigenvalues.
With each sequence obeying the statistics of the appropriate Dyson
ensemble one finds with Eq.~(\ref{eq:dyson})
\begin{eqnarray}
P(\{T_n\})
&\propto& \sum_{I\in {\cal P}}
\prod_{m>n} \left(T_{I_m}-T_{I_n}\right)^\beta
\prod_{m>n} \left(T_{\bar I_m}-T_{\bar I_n}\right)^\beta
\nonumber\\
&&\times \prod_{l=1}^N T_l^{-1+\beta/2}. \label{eq:cis}
\end{eqnarray}

For large $N$, the separation into two effectively independent
systems with Dirichlet and Neumann boundary conditions
naturally results in a doubling of the conductance fluctuations
(\ref{eq:varg}) and a doubling of the weak-localization
correction (\ref{eq:dg}). Moreover, level repulsion is only
effective for transmission eigenvalues which are part of the
same sequence. This modifies the spacing probability density,
which can be calculated from the general expression
~\cite{mehta-2004}
\begin{equation}
P(s)=\frac{d^2}{ds^2}\prod_i \int_0^\infty\int_0^\infty
p_i(\frac{\rho_i}{\rho}s+y+z)dydz
\end{equation}
for multiple sequences $i$, where $p_i(s)$ is the spacing
probability densities of each sequence, while
$\frac{\rho_i}{\rho}$  is the associated fractional eigenvalue density.

For two sequences following the Wigner distribution
(\ref{eqn:wignerdistribution}), the resulting spacing probability
densities is ($\bar s\equiv1$)
\begin{subequations}
\label{eq:wignersup}
\begin{eqnarray}
P_{\beta=1}(s)&=&\frac{e^{-2x^2}}{2}+\frac{\sqrt{\pi}}{2} x
e^{-x^2}\mathcal{E}(x),
\quad x= \frac{\sqrt{\pi} s}{4}\label{eq:wignersupa}
\\
P_{\beta=2}(s)&=&\frac{6x^2e^{-2x^2}}{\pi}
+2\frac{x-x^3}{\sqrt{\pi}}e^{-x^2}\mathcal{E}(x)+\frac{\mathcal{E}^2(x)}{2},
\nonumber\\&& \quad x=\frac{s}{\sqrt{\pi}} \label{eq:wignersup2a}
\\
P_{\beta=4}(s)&=&\frac{x}{3\sqrt{\pi}}(6+4
x^2-4x^4)e^{-x^2}\mathcal{E}(x)+\frac{\mathcal{E}^2(x)}{2} \nonumber \\
&&\hspace*{-1cm}{}+\frac{2 x^2}{9 \pi}(9+28x^2+8x^4)e^{-2x^2}
,\quad x=\frac{4s}{3\sqrt{\pi}},
 \label{eq:wignersup3a}
\end{eqnarray}
\end{subequations}
where $\mathcal{E}(x)=\mathrm{erfc}\,(x)$ denotes the
complementary error function.

\subsection{$\beta=1$: Reformulation as a staggered eigenvalue sequence}

In most situations encountered in random-matrix theory, the
combinatorial sum over partitions involved in the superposition
of eigenvalue sequences
([as in Eq.~(\ref{eq:cis})]
cannot be performed explicitly.
For the specific case $\beta=1$, however, the combinatorial sum
over $I$ in Eq.\ (\ref{eq:cis}) can be carried out (see below),
which then yields a closed-form expression
\begin{equation}P(\{T_n\})\propto\prod_{m>n,\atop \rm both\,odd}(T_m-T_n)
\prod_{m>n,\atop \rm
both\,even}(T_m-T_n)\prod_l\frac{1}{\sqrt{T_l}}.
 \label{eq:cis2}
\end{equation}
 (A similar simplification does
not present itself in the cases $\beta=2$ and $\beta=4$.) This
result finds its natural statistical interpretation as a {\em
staggered} superposition of two sequences, which is illustrated in
Fig.\ \ref{fig:leadpreserving}(d). In such a superposition, the
transmission eigenvalues in each sequence are not distinguished by the parity of the
associated wavefunction under the symmetry operation. Instead, the
transmission eigenvalues are ordered by magnitude (irrespective of parity), and one
sequence is composed of all odd-indexed transmission eigenvalues (of
which there are $N_1$) while the other sequence is composed of all
even-indexed transmission eigenvalues (of which there are $N_2$).
Compared to the original superposition of two independent
sequences, this differs by the additional constraint
\begin{equation}T_{I_1}\leq T_{\bar I_1}\leq
T_{I_2}\leq T_{\bar I_2}\leq T_{I_3}\leq T_{\bar I_3}\ldots
\label{eq:constraint}
\end{equation}
(which is satisfied when all the ordered indices $I_n$  are odd
while the indices $\bar I_n$ are all even).

 In order to
demonstrate the equivalence of Eq.\ (\ref{eq:cis}) (for $\beta=1$)
and Eq.\ (\ref{eq:cis2}) we have to show that the level-repulsion
terms are proportional to each other (both expressions share the
same product of one-point weights $\prod_lT_l^{-1/2}$, and the
proportionality constant is fixed by normalization). We set out to
work towards this goal by defining a matrix
\begin{equation}
M=\left(
\begin{array}{cccccccr}
-\mathbf{v}_1 & \mathbf{v}_2 & -\mathbf{v}_3 &\mathbf{v}_4
&-\mathbf{v}_5 &\mathbf{v}_6 &\ldots& (-1)^N\mathbf{v}_N
\\
 \mathbf{w}_1 &  \mathbf{w}_2 &  \mathbf{w}_3 &  \mathbf{w}_4 &
\mathbf{w}_5 & \mathbf{w}_6&\ldots & \mathbf{w}_N
\end{array}
\right), \label{eq:mmatrix}
\end{equation}
which is composed of column vectors
\begin{eqnarray}
&&\mathbf{v}_n=(1,T_n,T_n^2,\ldots,T_n^{N_1-1})^T,\quad
\\
&&\mathbf{w}_n=(1,T_n,T_n^2,\ldots,T_n^{N_2-1})^T.
\end{eqnarray}

The determinant $\det M$ can be evaluated in two different ways.
In the first way, we expand it in terms of subdeterminants with
$N_1$ vectors $\mathbf{v}_n$ from the first $N_1$ rows and $N_2$
vectors $\mathbf{w}_m$ from the remaining rows. In other words, we
sum over all determinants of the form
\begin{equation}
\det\left(
\begin{array}{ccccccc}
-\mathbf{v}_1 & \mathbf{v}_2 & 0 &\mathbf{v}_4 &0 &0 &\ldots
\\
0 & 0 & \mathbf{w}_3 & 0 & \mathbf{w}_5 & \mathbf{w}_6&\ldots
\end{array}
\right),
\end{equation}
etc., where the indices of the vectors ${\bf v}_{I_n}$ form an
ordered subsequence $I$ and the indices of the vectors ${\bf
w}_{\bar I_n}$ are given by the complementary subsequence $\bar
I$. The alternating signs in front of the vectors ${\bf
v}_{I_n}$ can be pulled out of the determinant at the cost of
an overall factor $(-1)^{I_1+I_2+\ldots+I_{N_1}}$. Next, we use
permutations of neighboring rows to bring all vectors
$\mathbf{v}_{I_n}$ to the left (into row $n$). This results in
an additional sign factor
$(-1)^{(I_1-1)+(I_2-2)+\ldots+(I_{N_1}-N_1)}$. The determinant
of the resulting block matrix factorizes.
Overall, this expansion yields
\begin{eqnarray}
&&\det M=(-1)^{N_1(N_1+1)/2} \nonumber \\ &&\times\sum_{I\in{\cal
P}}\det(\mathbf{v}_{I_1},\mathbf{v}_{I_2},\ldots,\mathbf{v}_{I_{N_1}})
\det(\mathbf{w}_{{\bar I}_1},\mathbf{w}_{{\bar
I}_2},\ldots,\mathbf{w}_{{\bar I}_{N_2}}). \nonumber \\
\label{eq:mexpansion1}
\end{eqnarray}
 Each
subdeterminant is of the form of a Vandermonde determinant, and
therefore
\begin{eqnarray}
\det M & = &(-1)^{N_1(N_1+1)/2} \nonumber \\ &&\times \sum_{I\in
{\cal P}}
\prod_{m>n} \left(T_{I_m}-T_{I_n}\right)
\prod_{m>n}
\left(T_{\bar I_m}-T_{\bar I_n}\right). \nonumber \\
\label{eq:detmres1}
\end{eqnarray}

Secondly, the determinant  $\det M$ can be evaluated by adding
in Eq.\ (\ref{eq:mmatrix}) the first $N_2$ rows to the last
$N_2$ rows. This yields
\begin{equation}
\det M=\det\left(
\begin{array}{ccccccc}
-\mathbf{v}_1 & \mathbf{v}_2 & -\mathbf{v}_3 &\mathbf{v}_4 &
-\mathbf{v}_5 & \mathbf{v}_6 &\ldots
\\
0 & 2\, \mathbf{w}_2 & 0 & 2\, \mathbf{w}_4 & 0 & 2\,
\mathbf{w}_6&\ldots
\end{array}
\right).
\end{equation}
Proceeding again with the evaluation of subdeterminants we are
left with a single choice, namely, to select vectors
$\mathbf{v}_n$ with odd index and  vectors $\mathbf{w}_n$ with
even index. Accounting for all signs and now also factors of
two, this results in
\begin{eqnarray}
\det M&=&(-1)^{N_1(N_1+1)/2} \,2^{N_2}\nonumber
\\ && \times
\det(\mathbf{v}_{1},\mathbf{v}_{3},\mathbf{v}_{5},\ldots)\det(\mathbf{w}_{2},\mathbf{w}_{4},\mathbf{w}_{6},\ldots).\qquad
\end{eqnarray}
As this again involves Vandermonde determinants, we find
\begin{eqnarray}
\det M &=&(-1)^{N_1(N_1+1)/2} \,2^{N_2} \nonumber
\\ && \times \prod_{m>n,\atop\rm
both\,odd}(T_m-T_n)\prod_{m>n,\atop\rm both\,even}(T_m-T_n).\qquad
\label{eq:detmres2}
\end{eqnarray}

The two results Eqs.\ (\ref{eq:detmres1}) and (\ref{eq:detmres2})
deliver the remarkable identity
\begin{eqnarray}
&&\sum_{I\in {\cal P}} \prod_{m>n}
\left(T_{I_m}-T_{I_n}\right)\prod_{m>n}
\left(T_{\bar I_m}-T_{\bar I_n}\right) \nonumber \\
&&=  2^{N_2}\prod_{m>n,\atop\rm
both\,odd}(T_m-T_n)\prod_{m>n,\atop\rm both\,even}(T_m-T_n).
\label{identity}
\end{eqnarray}
An equivalent identity has been derived for superpositions of energy
eigenvalue sequences with a length difference of at most one which are
distributed according to the Gaussian orthogonal ensemble \cite{forrester}.
Relation (\ref{identity}) shows that the level-repulsion term in Eq.\
(\ref{eq:cis}) is indeed proportional to the level-repulsion term in Eq.\
(\ref{eq:cis2}). As already mentioned, the one-point product
$\prod_lT_l^{-1/2}$ in both expressions is identical, and the proportionality
constant is fixed by normalization. It follows that for $\beta=1$, the
independent superposition of two transmission-eigenvalue sequences with $N_1$
and $N_2$ levels (with $N_1$ and $N_2$ constrained to differ at most by one)
is identical to a staggered superposition of two transmission-eigenvalue
sequences with $N_1$ and $N_2$ levels, which are correlated by the ordering
requirement (\ref{eq:constraint}).

\section{Lead-transposing symmetries\label{sec:3}}

\begin{figure}[t]
\includegraphics[width=0.8\columnwidth, keepaspectratio]{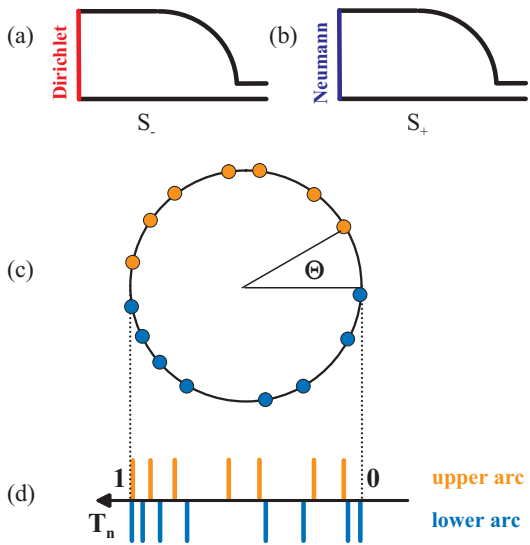}
\caption{(Color online) (a,b) Desymmetrization of the quantum
billiard with a lead-transposing reflection symmetry, shown in
Fig. \ref{fig1}(e). (c,d) Eigenphases $\Theta_n$ of the matrix $Q$
on the unit circle, and their projection Eq.\ (\ref{eq:taun})
which delivers the transmission eigenvalues.}
 \label{fig:leadtransposing}
\end{figure}

Systems with a lead-transposing symmetry require a separate
treatment since the symmetry operation only commutes with the
Hamiltonian, but not with the current operator (which changes its
sign). In the presence of an applied bias, the symmetry operation
exchanges the electronic source and drain reservoirs. An obvious
symptom of this complication is the fact that the desymmetrized
system only possesses a single lead (see Fig.\
\ref{fig:leadtransposing}). Mathematically, the transmission
matrix does not assume a block structure but remains full. We will
first adapt the concept of desymmetrization to derive the
constraints of the scattering matrix, and then turn to the joint
probability density of the transmission eigenvalues in
random-matrix theory. Just as in the previous section, we then
focus on the orthogonal symmetry class ($\beta=1$) and derive a
closed expression for the joint probability density, which again
assumes the form of a staggered level repulsion.

\subsection{Constraints on the scattering matrix}

The presence of a lead-transposing symmetry immediately results in
the constraint $r=r'$, $t=t'$ (when time-reversal symmetry is
broken by a magnetic field, this can be achieved by an inversion
symmetry but not by a reflection symmetry). In order to further
exploit the consequences of the symmetry, let us inspect a
time-reversal symmetric system with a reflection symmetry, as
shown in Fig.\ \ref{fig1}(e). As shown in Fig.\
\ref{fig:leadtransposing}(a,b), the desymmetrized versions are cut
at the symmetry line, where they are equipped with Dirichlet or
Neumann boundary conditions for wavefunctions of odd ($-$) or even
parity ($+$), respectively. Such wave functions are readily
constructed starting from the original system when one chooses
incoming amplitudes of the form ${\bf a}^{(R)}=\pm {\bf a}^{(L)}$.
The outgoing amplitudes are then given by ${\bf b}^{(L)}=(r\pm
t){\bf a}^{(L)}$. Consequently, the scattering matrices of the
desymmetrized systems are given by
\begin{equation} S_{\pm}=r\pm t. \label{eq:spm}
\end{equation}

The desymmetrized systems only possess a single opening. In order
to revert to the scattering matrix of the original system we
invert Eq.\ (\ref{eq:spm}). The transport in the original system
is therefore described by the transmission matrix
$t=\frac{1}{2}(S_+-S_-)$, which gives
\begin{equation}\label{ttdagger}
tt^\dagger=\frac{1}{4}(2-S_+S_-^\dagger-S_-S_+^\dagger).
\end{equation}
The properties of this matrix---and especially, of its
eigenvalues $T_n$---are not separable and depend on the
interplay of both desymmetrized variants.

\subsection{Conventional random-matrix theory\label{sec:3conv}}

Random-matrix ensembles for systems with lead-transposing symmetry
can be obtained by assuming that the scattering matrices $S_+$ and
$S_-$  of the desymmetrized variants are statistically independent
realizations of the appropriate standard circular ensemble. The
resulting ensembles are identical to those introduced by Baranger
and Mello \cite{BM96}, who based their considerations on a
maximal-entropy principle.

Earlier works have addressed isolated aspects of these ensembles,
but not the complete transmission-eigenvalue statistics. For
instance, it has been observed that a lead-transposing symmetry
increases the conductance fluctuations \cite{GMMB96,gopar-2006-73}
but eliminates the weak-localization correction
\cite{gopar-2006-73}. For large $N$, the conductance fluctuations
double, just as is the case for lead-preserving symmetries. We now
provide a complete explanation of these observations on the basis
of the joint probability density of the transmission eigenvalues.

The starting point of these considerations is the relation
\begin{equation} \label{eq:taun}
T_n=\sin^2(\Theta_n/2)=\frac{1}{2}\left(1-\cos\Theta_n\right)
\end{equation}
between the transmission eigenvalues $T_n$ and the eigenphases
$\Theta_n$ of the unitary matrix $Q\equiv S_+S_-^\dagger$, which
follows from Eq.\ (\ref{ttdagger}). As illustrated in Fig.\
\ref{fig:leadtransposing}(c,d), the statistics of transmission
eigenvalues is hence directly imposed by the statistics of the
real parts $\cos\Theta_n$ of the unimodular eigenvalues
$e^{i\Theta_n}$ of $Q$.

In random-matrix theory, the eigenphases $\Theta_n$ follow the
statistics of the associated circular ensemble. This is evident
for the unitary ensemble ($\beta=2$), which is invariant under the
multiplication of an arbitrary fixed matrix (it hence suffices,
e.g., to assume that $S_+$ is random while $S_-$ is fixed, or vice
versa). In the orthogonal case ($\beta=1$), the unitary
transformation $Q'=S_-^{-1/2}QS_-^{1/2}=S_-^{-1/2}S_+S_-^{-1/2}$
results in a symmetric matrix with identical eigenvalues. Their
circular statistics then follows from the fact that the circular
orthogonal ensemble is invariant under the symmetric involution
with any fixed symmetric matrix (here, $S_-^{-1/2}$). The same
transformation also succeeds in the case of self-dual matrices
($\beta=4$).

Because of the uniform distribution of eigenphases in the circular
ensemble \cite{mehta-2004}, the one-point probability density
$P(T_n)$ is given by Eq.\ (\ref{eq:onepoint}) for any finite $N$
(i.e., not only in the limit $N\to\infty$) \cite{gopar-2006-73}.
The joint probability  density of the eigenphases $\Theta_n$ is
given by
 \cite{mehta-2004}
\begin{equation}
P_\Theta(\{\Theta_n\})\propto\prod_{m>n}\left[\sigma_m\sin\frac{\Theta_m-\Theta_n}{2}\right]^\beta.
\label{eq:theta1}
\end{equation}
Here we ordered the eigenphases by their moduli,
\begin{equation}
0\leq|\Theta_1|\leq|\Theta_2|\leq|\Theta_3|\leq\ldots\leq|\Theta_N|\leq\pi,
\label{eq:thetaordering}
\end{equation}
and denoted $\sigma_n={\rm sgn}\,\Theta_n$. Since Eq.\
(\ref{eq:taun}) does not discriminate the sign of $\Theta_n$ we
proceed to the distribution of the moduli $\theta_n=|\Theta_n|$,
\begin{equation}
P_\theta(\{\theta_n\})=\sum_{\{\sigma_n\}}P_\Theta(\{\sigma_n\theta_n\}).
\label{eq:theta2}
\end{equation}
With the help of the relations
\begin{equation}
\sin(\theta_n/2)=\sqrt{T_n},\quad \cos(\theta_n/2)=\sqrt{1-T_n},
\label{eq:trafo1}
\end{equation}
and also accounting for the Jacobian
\begin{equation}
\frac{d\theta_n}{d T_n}=\frac{1}{\sqrt{T_n(1-T_n)}},
\label{eq:jacobian}
\end{equation} this
yields the joint probability density \cite{gopar-2006-73}
\begin{eqnarray}
&&
P(\{T_n\})
\propto \prod_{l}\frac{1}{\sqrt{T_l(1-T_l)}} \nonumber \\
&&\times\sum_{\{\sigma_n\}} \prod_{m>n}\left[
\sqrt{T_n(1-T_m)}-\sigma_m\sigma_n\sqrt{T_m(1-T_n)} \right]^\beta
.\nonumber \\
\label{eq:pjointtransposed}
\end{eqnarray}
This expression is symmetric under the replacement $T_n\to 1-T_n$,
which explains the absence of weak-localization corrections to the
conductance. Moreover, transmission eigenvalues do not repel each
other when $\sigma_n=-\sigma_m$, i.e., when the underlying
eigenphases $\Theta_n$ lie on the opposite (upper and lower) arcs
of the unit circle [see again Fig.\ \ref{fig:leadtransposing}(c)].
As the sets of eigenphases on both arcs is only weakly cross-correlated,
this explains the doubling of the conductance fluctuations for
large $N$.

\subsection{Staggered level repulsion for $\beta=1$}

While the general conclusions of the previous section can be drawn
for any $\beta$, it should be noted that Eq.\
(\ref{eq:pjointtransposed}) still involves a combinatorial sum,
and hence is similar in status as expression (\ref{eq:cis}) for
systems with a lead-preserving symmetry. We now
show that a much more detailed insight is possible for the
orthogonal symmetry class ($\beta=1$), where the combinatorial sum
in Eq.\ (\ref{eq:pjointtransposed}) can be carried out explicitly
(see below). The resulting statistics again assume the form of a
staggered level repulsion, but are not identical to Eq.\
(\ref{eq:cis2}) (which was derived from the superposition  of two
independent level sequences): For $N$ an odd integer, we find
\begin{subequations}\label{eq:result}%
\begin{eqnarray}
 P(\{T_n\})
&\propto& \prod_{m>n,\,\rm both\,odd}\!\!\!\!\!\!\!\!(T_m-T_n)
\prod_{l \,{\rm\, odd}}\frac{1}{\sqrt{T_l(1-T_l)}}  \nonumber \\
& \times & \prod_{m>n,\,\rm both\,even}\!\!\!\!\!\!\!\!(T_m-T_n)
\label{eq:result1}
,
\end{eqnarray}%
while for even $N$
\begin{eqnarray} P(\{T_n\})
& \propto & \prod_{m>n,\,\rm both\,odd}\!\!\!\!\!\!\!\!(T_m-T_n)
\prod_{l \,{\rm\, odd}}\frac{1}{\sqrt{T_l}} \nonumber
\\
& \times & \prod_{m>n,\,\rm both\,even}\!\!\!\!\!\!\!\!(T_m-T_n)
 \prod_{l\,{\rm\,even}}\frac{1}{\sqrt{1-T_l}}
 \label{eq:result2}
 .\qquad
\end{eqnarray}
\end{subequations}%
Similar to Eq.~(\ref{eq:cis2}), the joint probability density again separates
into two factors, each involving only every second eigenvalue. In particular,
neighboring levels are not prohibited to approach each other closely, and
statistical fluctuations of observables are enhanced, as has been earlier
observed for the conductance and the Fano factor
\cite{GMMB96,BM96,SSML05,gopar-2006-73}. The correlation between the two
level sequences is again imposed only indirectly by the requirement that the
sequences are staggered. This ordering requirement is independent of the
parity of the wave function -- indeed, in the present case, parity is not
well defined as the transmission eigenvalues arise from the combined
properties of $S_+$ and $S_-$.

In order to derive Eq.\ (\ref{eq:result}),
let us first inspect Eq.\ (\ref{eq:theta1}). Because of the
ordering (\ref{eq:thetaordering}), each factor $\sigma_m$ appears
$m-1$ times, and therefore
\begin{equation}
P_\Theta(\{\Theta_n\})\propto\prod_{l\,\rm
even}\sigma_l\prod_{m>n}\sin\frac{\Theta_m-\Theta_n}{2}.
\label{eqn:pTheta}
\end{equation}

We next pass over to the joint distribution of moduli
(\ref{eq:theta2}). In order to evaluate the combinatorial sum over
the $\sigma_n$ we express the factor of sine functions in Eq.\
(\ref{eqn:pTheta}) as a Vandermonde determinant,
\begin{equation}
\prod_{m>n}\sin\frac{\Theta_m-\Theta_n}{2}=(-i)^{N(N-1)/2}\det
B(\{\sigma_n\theta_n\}),
\end{equation}
where $B_{ml}(\{\Theta_n\})=\exp(i\Theta_ml)$, $m=1,2,3,\ldots,N$,
while the index $l$ runs in integer steps from $-(N-1)/2$ to
$(N-1)/2$.  The multilinearity of the determinant then yields
\begin{equation}
P_\theta(\{\theta_n\})\propto(-i/2)^{N(N-1)/2}\,{\rm det}\, C,
\end{equation}
where $C_{ml}=2\cos(\theta_m l)$ for odd $m$ and
$C_{ml}=2i\sin(\theta_m l)$ for even $m$.

For every $l>0$ we now add the $l$th column in $C$ to the $-l$th
column, which cancels all sine terms in the latter columns. The
determinant  $\det C=\det D \det E$ then factorizes, where
$D_{ml}=\cos{\theta_m l}$, $m$ odd, and $E_{ml}=\sin{\theta_m l}$,
$m$ even. If $N$ is even, the index $l$ is now restricted to
$l=1/2,3/2,\ldots,(N-1)/2$. For odd $N$, this index is restricted
to $l=0,1,2,\ldots,(N-1)/2$ for the matrix $D$, and to
$l=1,2,\ldots,(N-1)/2$ for the matrix $E$.

For odd $N$ we can write $D_{ml}$ as a polynomial of degree $l$ in
$\cos\theta_m$, and $E_{ml}$ as $\sin \theta_m$ times a polynomial
of degree $l-1$ in $\cos\theta_m$. We only need to keep the
highest monomial, as the other terms are linear combinations of
the rows of lower index $l$. This leaves us again with Vandermonde
determinants,
\begin{eqnarray}
\det D&\propto&\!\!\prod_{m>n,{\rm\,both\,
odd}}\!\!\!\!\!\!\!\!(\cos\theta_n-\cos\theta_m),
\\
\det E&\propto&\prod_{l{\,\rm\,
even}}\sin\theta_l\!\!\!\!\!\!\!\prod_{
m>n,{\rm\,both\,even}}\!\!\!\!\!\!\!\!(\cos\theta_n-\cos\theta_m).
\end{eqnarray}

For even $N$, the index $l$  is half-integer, and the elements of
$D$ can now be written as $\cos(\theta_m/2)$ times a polynomial in
$\cos(\theta_m)$, while those of $E$ can be written as
$\sin(\theta_m/2)$ times such a polynomial. This yields
\begin{eqnarray}
\det D&\propto&\prod_{l{\,\rm\,
odd}}\cos(\theta_l/2)\!\!\!\!\!\!\!\prod_{m>n,{\rm\,both\,
odd}}\!\!\!\!\!\!\!\!(\cos\theta_n-\cos\theta_m),\quad
\\
\det E&\propto&\prod_{l{\,\rm\,
even}}\sin(\theta_l/2)\!\!\!\!\!\!\!\!\prod_{
m>n,{\rm\,both\,even}}\!\!\!\!\!\!\!\!\!
(\cos\theta_n-\cos\theta_m).\quad
\end{eqnarray}
The joint probability density (\ref{eq:result}) follows by
transforming from $\theta_n$ to $T_n$, where the Jacobian is given
by Eq.\ (\ref{eq:jacobian}), while the factors in the expressions
for $D$ and $E$ can be rewritten with the help of Eq.\
(\ref{eq:trafo1}) and the relation
$\cos\theta_n-\cos\theta_m=2(T_m-T_n)$.

\subsection{Large-$N$ asymptotics\label{sec:3a}}

It is natural to ask whether the similarity of Eq.\
(\ref{eq:result}) to Eq.\ (\ref{eq:cis2}) indicates a possible
interpretation as a superposition of two independent level
sequences [from which Eq.\ (\ref{eq:cis2}) was derived]. In Eq.\
(\ref{eq:result}), however, this interpretation is prevented by
the different one-point weight terms associated to the even and
odd indexed eigenvalues. A symptom of this difference is the fact
that Eq.\ (\ref{eq:cis2}) implies finite-$N$ weak-localization
corrections to the conductance, while Eq.\ (\ref{eq:result})
delivers the absence of such corrections, in agreement with the
general conclusions in Sec.\ \ref{sec:3conv}. Hence, the
statistics of systems with a lead-transposing and a
lead-preserving symmetry (with $\beta=1$)  only find a common
ground when both are interpreted as a staggered level sequence.

For the case of a lead-preserving symmetry, the framework of
superpositions of independent level sequences of course provides a
powerful tool for the derivation of low-point correlation
functions and local statistics [such as the two-point correlation
function, or the level spacing distribution (\ref{eq:wignersup})].
We now argue that in the limit $N\to \infty$, this framework can
also be adopted for systems with a lead-transposing symmetry.

In this limit, the transmission eigenvalues form a
quasi-continuum, and the asymptotical statistics follow from the
formal analogy to the statistics of coordinates of a dense set of
parallel line charges in one dimension (the Coulomb gas), which
exhibit a logarithmic repulsion \cite{mehta-2004,B97}. In leading
order, the weight terms enter the analysis of the statistical
fluctuations only via the one-point function $P(T)$: For fixed
index $n$, the transmission eigenvalue $T_n$ are confined to a
small neighborhood around a nominal equilibrium position $\bar
T_n$, which is given by the implicit equation $n-1/2=N\int_0^{\bar
T_n} P(T) dT$. Subsequently, the weight terms can be approximated
by a constant (with all the $T_n$ fixed to $\bar T_n$), while  the
fluctuations are exclusively governed by the level-repulsion
factors of the joint probability distribution. As the
level-repulsion factors are identical in Eqs.\ (\ref{eq:cis2}) and
(\ref{eq:result}) one concludes that the local statistics in both
ensembles become indistinguishable in the limit of $N\to\infty$.

We therefore obtain the following  remarkable result of purely
statistical origin: For a lead-transposing symmetry, as $N$ is
sent to infinity  the local statistics (embodied in low-point
correlation functions) converges to that of a superposition of two
independent level sequences. This is the case even though a
classification of transmission eigenvalues by parity is not
possible. In particular, we arrive at the prediction that in this
limit, the level-spacing distribution is well approximated by Eq.\
(\ref{eq:wignersup}).

\section{Numerical investigations\label{sec:4}}

For the three standard Dyson ensembles of random-matrix theory,
the joint probability density (\ref{eq:dyson})  manifests the
celebrated repulsion between neighboring eigenvalues, since the
probability to find two closely spaced adjacent eigenvalues is
suppressed as $(T_{n+1}-T_{n})^\beta$. In contrast, the joint
densities (\ref{eq:cis2}) and (\ref{eq:result}) (both derived for
$\beta=1$) describe sequences of reduced stiffness, where only
every second level is subject to mutual level repulsion. As argued
before, as long as $N$ takes on moderate values, the latter joint
densities imply quantitative differences in the transmission
eigenvalue statistics for lead-preserving and lead-transposing
symmetries, while for large $N$ these statistics should converge
onto each other.

In this section we illustrate the differences and similarities
between these scenarios for all three main symmetry classes
($\beta=1,2,4$)  via numerical sampling of the random-matrix
ensembles, and also compare to realistic model systems. For
convenient characterization of the eigenvalue repulsion we employ
the nearest-neighbor spacing distribution $P(s)$, as well as
spacing distributions to more distant neighbors. As we will see,
the local statistics of systems with a lead-transposing symmetry
actually show a much {\em weaker} $N$ dependence than for systems
with a lead-preserving symmetry. This feature could be anticipated
by (but also goes beyond) the absence of weak localization
corrections in the one-point function (discussed in Sec.\
\ref{sec:3conv}).

\subsection{Random-matrix theory}

\begin{figure}
\includegraphics[width=0.9\columnwidth, keepaspectratio]{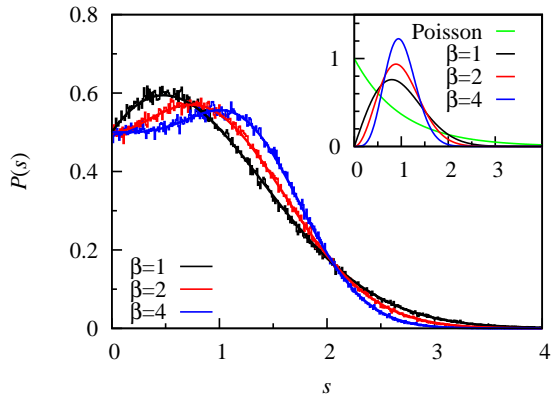}
\caption{(Color online) Probability density $P(s)$ of
transmission-eigenvalue spacings for systems with a
lead-transposing symmetry, obtained from $10^4$ random matrices
with $N=50$. Smooth curves: Spacing probability density (\ref{eq:wignersup}) for
superpositions of eigenvalues of two independent sequences from
the standard circular ensembles. The inset shows the Wigner
distributions (\ref{eqn:wignerdistribution}) from standard
random-matrix theory and the Poisson distribution
(\ref{eq:poisson}) for uncorrelated eigenvalues. \label{fig2}}
\end{figure}

\begin{figure}
\includegraphics[width=\columnwidth, keepaspectratio]{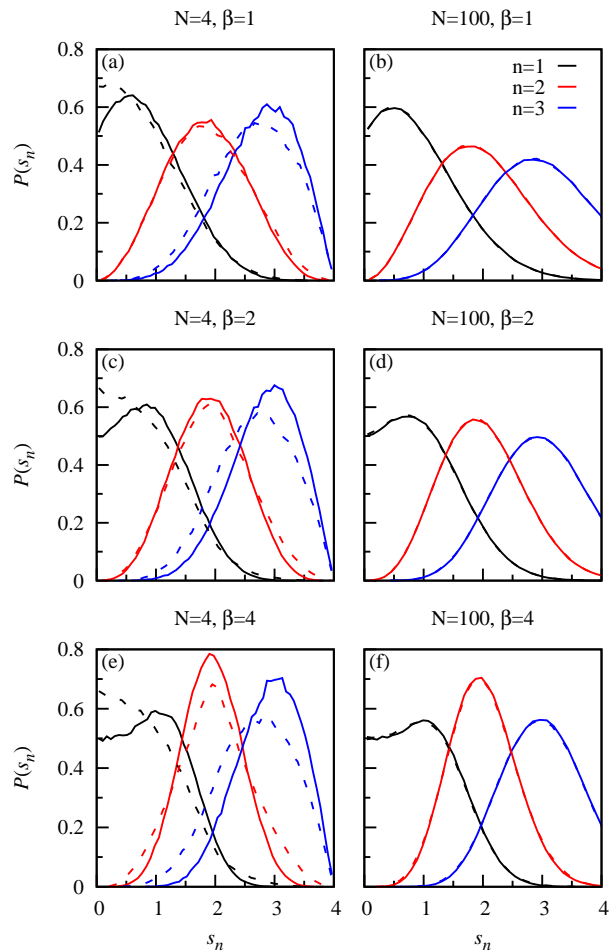}
\caption{(Color online) Probability densities of spacings $s_n$ to
the first, second and third neighboring transmission eigenvalue
for the random-matrix ensembles of systems with a lead-transposing
symmetry (solid curves) or a lead-preserving symmetry (dashed
curves). In the left panels the number of transport channels
$N=4$, while in the right panels $N=100$. Top panels: orthogonal
symmetry class ($\beta=1$). Middle panels: unitary symmetry class
($\beta=2$). Bottom panels: symplectic symmetry class ($\beta=4$).
For each ensemble, the results represent a sample of $10^4$
realizations. \label{fig4}}
\end{figure}

We start with the characterization of the level statistics within
the various random-matrix ensembles. Let us first consider the
case of a lead-transposing symmetry with a relatively large
number
of transport channels,  for which we expect that the local
statistics is close to that of a superposition of two independent
level sequences. Starting point of the numerical computations is
Eq.\ (\ref{ttdagger}), where the matrix $Q=S_+S_-^\dagger$ is
drawn from the appropriate Dyson ensemble. In order to obtain the
nearest-neighbor spacing distribution $P(s)$, we unfold the
eigenvalue sequences to a mean local spacing $\bar s\equiv 1$
\cite{mehta-2004,haake-2001}. Figure \ref{fig2} shows the
resulting spacing distributions for $N=50$. For this large number
of open channels we find that the numerical histograms indeed
match the predictions from the superposition of two independent
level sequences [solid curves;
see Eq.\ (\ref{eq:wignersup})].

For comparison, the inset in Fig.\ \ref{fig2} shows the standard
Wigner distributions (\ref{eqn:wignerdistribution}), as well as
the Poisson distribution (\ref{eq:poisson}). In the Poisson
distribution  the eigenvalue spacing density is maximal at $s=0$;
for larger $s$ the probability density decreases monotonically.
For the Wigner distributions the most likely eigenvalue spacing
occurs at a finite value of $s$; for $s\to 0$, the distributions
decay algebraically $\propto s^\beta$, while for $s\to\infty$ they
decay as a Gaussian. The distributions in the main panel combine
the partial absence of level repulsion for small $s$ [with
$P(s=0)=1/2$] with the Gaussian decay of the Wigner distributions
for large $s$.

For large $N$, virtually identical results are obtained for the
conventional case of a lead-preserving symmetry. This is
demonstrated in detail in Fig.\ \ref{fig4}, which also shows the
spacing distributions to the second and third-nearest neighbor.
Here, solid curves are for a lead-transposing symmetry, and dashed
curves are for a lead-preserving symmetry (corresponding to a
superposition of independent level sequences from the appropriate
Dyson ensemble). For $N=100$ (right panels), dashed and solid
curves lie on top of each other and are practically
indistinguishable. This clearly supports the convergence of the
local statistics of both cases for large $N$.

The left panels  in Fig.\ \ref{fig4} show the level-spacing
distributions for $N=4$. In this case, the results for a
lead-transposing symmetry are distinctively different from those
for a lead-preserving symmetry. Interestingly, the
nearest-neighbor spacing distribution for a lead-transposing
symmetry
is very similar for small and large $N$;
the
distribution for $N=4$ is already well approximated by Eq.\
(\ref{eq:wignersup}). In comparison, the nearest-neighbor spacing
distribution for a lead-preserving symmetry shows a much stronger
$N$-dependence.

\subsection{Comparison to model systems}

\begin{figure}
\includegraphics[width=\columnwidth, keepaspectratio]{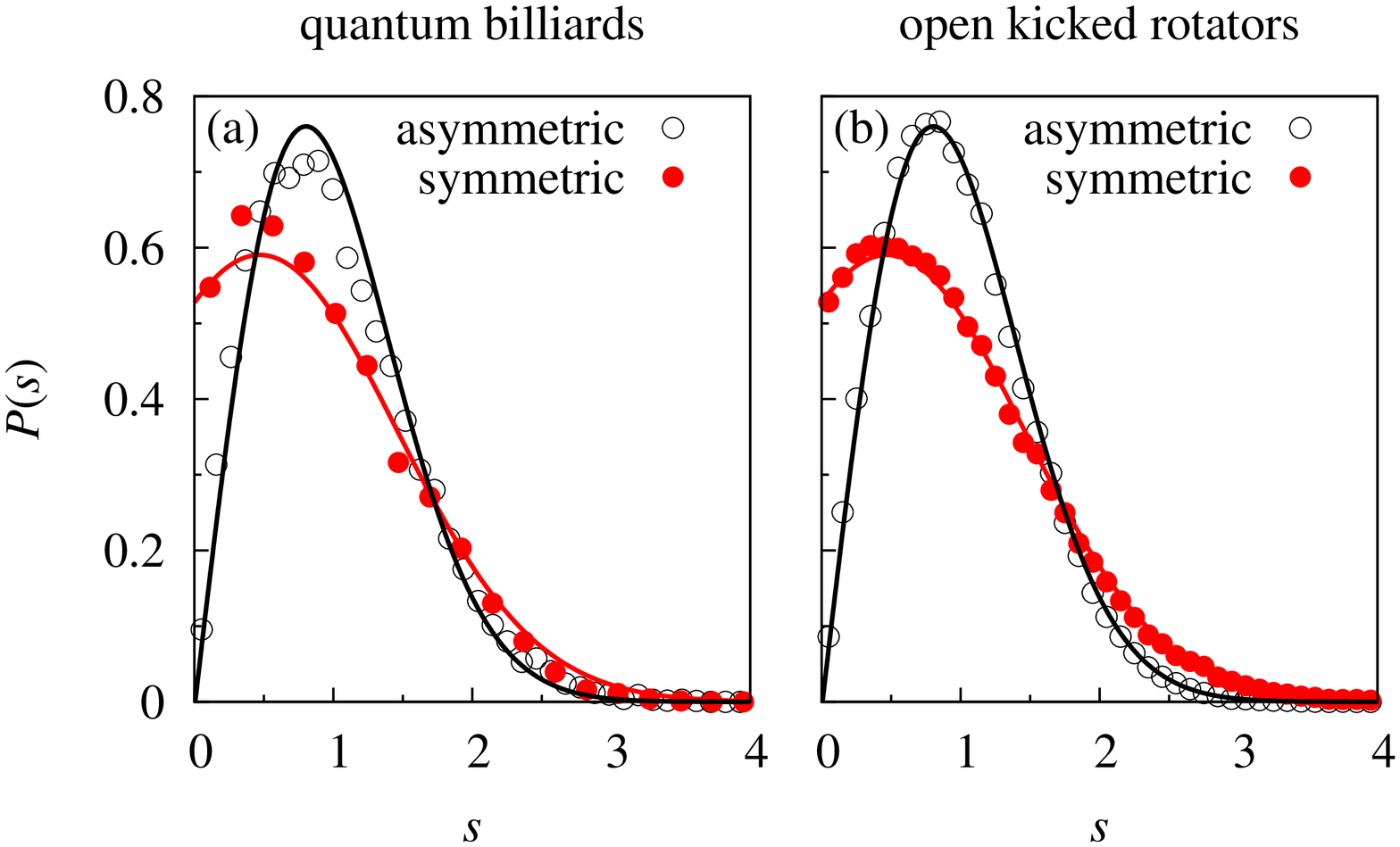}
\caption{(Color online) (a) Nearest-neighbor spacing distribution
$P(s)$ for the lead-asymmetric stadium billiard of Fig.
\ref{fig1}(b), averaged over energies in the range $N=5-14$, and
the lead-transposing symmetric stadium billiard of Fig.\
\ref{fig1}(e), with $N=5,6$. (b) The same for open quantum kicked
rotators with $N=12$. In both panels, the solid curves show the
Wigner distribution (\ref{eqn:wignerdistribution}) with $\beta=1$
and the prediction of random-matrix theory for systems with a
lead-transposing symmetry [which can be safely approximated by
Eq.\ (\ref{eq:wignersupa})]. \label{fig3}}
\end{figure}

In order to validate
that realistic quantum systems can indeed be described by random
matrix theory (on which all previous considerations are based), we
compare our predictions with numerical results for such systems.
In particular, we present
results of numerical computations for
quantum billiards, which model a lateral quantum dot, and
for
the open
kicked rotator, which is based on an efficient quantum map. We
focus on systems in the orthogonal symmetry class ($\beta=1$) and
contrast systems with a lead-transposing symmetry to systems
without any spatial symmetry.

The quantum billiards are derived from the stadium geometry, with
leads positioned to either break or conserve the reflection
symmetry about the vertical center line [see Figs.\
\ref{fig1}(b,e)]. The computations are performed using a modular
recursive Green's function method \cite{RTWTB00,RAB07}, with
energies that permit $5\leq N\leq 14$ open channels in each of the
two leads. As shown in Fig.\ \ref{fig3}(a), the eigenvalue spacing
distribution agrees well with the predictions of random-matrix
theory, both in presence and in absence of a lead-transposing
symmetry.

The open quantum kicked rotator \cite{TTSB03, JSB03, OKG03} is
defined by the scattering matrix
\begin{equation}
S=P[e^{-i\varepsilon}-F(1-P^T P)]^{-1}FP^T,
\end{equation}
where $\varepsilon$ is the quasi-energy,
\begin{equation}
F_{nm}=(iM)^{-1/2}
e^{\frac{i\pi}{M}(m-n)^2-\frac{iMK}{4\pi}(\cos\frac{2\pi
n}{M}+\cos\frac{2\pi m}{M})}
\end{equation}
is the $M\times M$-dimensional Floquet operator of the kicked
rotator, and $P$ is an $2N\times M$-dimensional matrix which
projects the internal Hilbert space onto the openings.  We assume
that $M$ is even and $M\gg N$. The reflection symmetry of the
closed system is manifested in the symmetry $F_{nm}=F_{M-n,M-m}$,
and the lead-transposing symmetry of the open system is present
when in addition $P_{nm}=P_{2N-n,M-m}$.

Figure \ref{fig3}(b) shows the spacing distributions obtained for
kicked rotators with symmetrical and asymmetrical lead placement
and $N=12$. The data  represents  6600 realizations which are
generated by varying the quasienergy $\varepsilon\in [0,2\pi)$,
the kicking strength $K\in[10,15]$, and the internal dimension
$M\in[498,502]$. Again, we find good agreement with random-matrix
theory, including the reduced eigenvalue repulsion in the
lead-symmetric case.

The results in this section reveal clear signatures of staggered
level repulsion in realistic systems with a lead-transposing
symmetry (and $\beta=1$). It is worth emphasizing that the
applicability of this statistical concept [embodied in the
random-matrix results Eq.\ (\ref{eq:result})] does not rely on any
pre- or postprocessing or -selection of the transmission
eigenvalues in the model systems (as there is no intrinsic
property of the transmission eigenvalues or their associated
scattering wave functions -- such as a parity -- that could be
used to divide these eigenvalues into two sets).

\section{Summary and Conclusions\label{sec:5}}

We analyzed the transport in open systems with a lead-transposing
or a lead-preserving symmetry via the complete joint probability
density of transmission eigenvalues, obtained in random-matrix
theory.

For a lead-preserving symmetry, the standard concept of
desymmetrization reduces the problem to the investigation of
independent non-symmetric  variants of the system. For a
lead-transposing symmetry, however, the transport characteristics
only arise as a collective property of the symmetry-reduced
variants of the system. We still found that both types of symmetry
result in a similar reduction of level repulsion, so that
transmission eigenvalues can approach each other closely. For a
large number of transport channels $N$, the local eigenvalue
statistics for both types of symmetry indeed become
indistinguishable.

Our main analytical results concern a detailed explanation of
these features for  systems which also exhibit time-reversal and
spin-rotation invariance (the orthogonal symmetry class, with
symmetry index $\beta=1$). In this case, the transmission
eigenvalue statistics of systems with a lead-transposing or
lead-preserving symmetry find a common natural interpretation as a
staggered superposition of two independent level sequences. In
such a superposition the eigenvalues alternate between the
sequences when they are ordered by magnitude. The joint
probability densities for the two types of symmetry only differ in
one-point weight factors. For lead-preserving symmetries these
weight factor incorporate $1/N$ corrections for quantities such as
the ensemble-averaged conductance, while these corrections are
absent for a lead-transposing symmetry. This results in
differences of the local eigenvalue statistics when $N$ is small,
but becomes insignificant when $N$ is large.

While we concentrated on systems with discrete spatial symmetries,
our results can also be  applied for discrete symmetries of
different origin (e.g., arising from internal degrees of freedom)
that yield equivalent constraints on the scattering matrix.

\begin{acknowledgments}
We gratefully acknowledge assistance with the billiard
computations by Florian Aigner, as well as useful discussions with
Eugene Bogomolny, Piet Brouwer, Victor Gopar, Jon Keating, and
Martin Zirnbauer. This work was supported by the European
Commission, Marie Curie Excellence Grant MEXT-CT-2005-023778. S.R.
wishes to thank the Max-Kade foundation and the W.M. Keck
foundation for support.
\end{acknowledgments}

\end{document}